\documentclass[preprint,showpacs,titlepage,aps,prd,
tightenlines,
amsmath,byrevtex,nofootinbib]{revtex4}

\usepackage{graphicx}

\def\lsim{\raise0.3ex\hbox{$\;<$\kern-0.75em\raise-1.1ex
\hbox{$\sim\;$}}}
\def\gsim{\raise0.3ex\hbox{$\;>$\kern-0.75em\raise-1.1ex
\hbox{$\sim\;$}}}

\begin{document}

\title{Correlated, Precision Measurements of $\theta_{23}$ and $\delta$ using only the Electron Neutrino Appearance Experiments}

\vskip 16mm 

\author{Hisakazu~Minakata}
\email{hisakazu.minakata@gmail.com}
\author{Stephen~J.~Parke}
\email{parke@fnal.gov} 
\affiliation{
Theoretical Physics Department, Fermilab, P.\ O.\ Box 500, Batavia, IL 60510, USA \\
}

\date{March 25, 2013}

\vglue 1.6cm

\begin{abstract}
Precision measurement of the leptonic CP violating phase $\delta$ will suffer from the, then surviving, large uncertainty of $\sin^2 \theta_{23}$ of $10-20\%$ in the experimentally interesting region near maximal mixing of $\theta_{23}$. We advocate a new method for determination of both $\theta_{23}$ and $\delta$ at the same time using only the $\nu_e$ and $\bar{\nu}_e$ appearance channels, and show that $\sin^2  \theta_{23}$ can be determined automatically with much higher accuracy, approximately a factor of six, than $\sin \delta$. In this method, we identify a new degeneracy for the simultaneous determination of $\theta_{23}$ and $\delta$, the $\theta_{23}$ intrinsic degeneracy, which must be resolved in order to achieve precision measurement of these two parameters. Spectral information around the vacuum oscillation maxima is shown to be the best way to resolve this degeneracy.

\end{abstract}


\maketitle

\section{Introduction}

In neutrino oscillation physics, the traditional way of determining the leptonic CP violating phase, $\delta$, is to measure $\delta$ together with $\theta_{13}$ by using $\nu_e$ and $\bar{\nu}_e$ appearance channels in muon neutrino superbeam  experiments, (or similarly by $\nu_\mu$ and $\bar{\nu}_\mu$ appearance in either neutrino factory or beta beam experiments). In posing the problem in this way, $\theta_{23}$ is assumed to be determined to high accuracy by a $\nu_\mu$ disappearance measurement up to the octant degeneracy. But, now we need a new orientation to choose the right approach for the determination of $\delta$, because: 
\begin{itemize}
\item $\theta_{13}$ is already determined with high accuracy, and its precision will become even greater by the time $\delta$ is measured \cite{Daya-Bay,RENO,DChooz,T2K,MINOS},
\item and $\theta_{23}$ will be determined with less percentage accuracy than $\theta_{13}$ in the experimentally preferred region,  see \cite{SK-atm, MINOS-disapp}, between $40^\circ$ and $50^\circ$. Since the disappearance measurements will have difficulty in determining $s^2_{23}\equiv \sin^2 \theta_{23}$ with an accuracy better than $10-20\%$  in this region \cite{MSS04}.
\end{itemize}
 
How seriously does the uncertainty in $s^2_{23}$ affect the sensitivity to $\delta$? We show in Appendix~\ref{delta-error} that the uncertainty of $ \delta$ is approximately proportional to
\begin{eqnarray}
 \sqrt{ (\Delta P/P)^2 + (\Delta  \bar{P}/\bar{P})^2 + \eta (\Delta s^2_{23}/s^2_{23})^2 }
\label{error-formula}
\end{eqnarray}
where $P$ and $\bar{P}$ are the $\nu_e$ and $\bar{\nu}_e$ appearance probabilities with uncertainties $ \Delta P$ and $\Delta  \bar{P}$, respectively, and $\Delta s^2_{23}$ is the uncertainty on $s^2_{23}$, and $\eta$ is generically a number of order unity. 
 Thus, when the measurement enters into a precision era in which the percentage uncertainty on $P$ and $\bar{P}$ would be smaller than, say 5\%, the uncertainty of $ \delta$ would be dominated by the $s^2_{23}$ uncertainty\footnote{
Note that this is a simplified discussion for the sake of clarity. A more detailed and accurate description can be found in Appendix A.
}.

In this paper, we present a new strategy for precision measurement of $\delta$ and $\theta_{23}$ to overcome this problem. Instead of using $\nu_\mu$ disappearance channel for determination of $\theta_{23}$, we rely on $\nu_e$ and $\bar{\nu}_e$ appearance measurement to determine precisely $\theta_{23}$ and $\delta$ at the same time. With this method the uncertainties expected for $\theta_{23}$ and $\delta$ are related with each other as follows
\begin{eqnarray}
\Delta (s^2_{23}) \simeq \frac{1}{6} \Delta (\sin \delta), 
\label{error-relation}
\end{eqnarray}
near the first vacuum oscillation maximum, as shown in Sec.~\ref{error-23-delta}. Therefore, once we enter into the era in which $\sin \delta$ can be measured with reasonably high accuracy, $s^2_{23}$ can be determined automatically with significantly higher accuracy than $\sin \delta$ which can be competitive with the disappearance measurement in the region $40^\circ < \theta_{23} < 50^\circ$.  Of course, the disappearance measurement will provide supplemental information, and also serves as a consistency check on $\nu$SM.

This new setting, of simultaneous measurement of $\theta_{23}$ and $\delta$, brings us to the problem of a new degeneracy involving these two parameters which we call the ``$\theta_{23}$ intrinsic degeneracy'' because of it's similarity to the $\theta_{13}$ intrinsic degeneracy \cite{intrinsic}. Notice that this degeneracy is entirely different from the $\theta_{23}$ octant degeneracy discussed in \cite{octant}.   We show that resolving this  $\theta_{23}$ intrinsic degeneracy is crucial to achieve precision measurements of $\delta$ and $\theta_{23}$, and that the disappearance measurement cannot help in region $\sin^2 2\theta_{23}  \gsim 0.97$. We argue that spectrum information around the vacuum oscillation maximum peaks is the most powerful way to resolve the degeneracy. It may be possible that such spectrum analyses could be carried out with a narrow-band neutrino beam, but if it is not powerful enough, use of a wide-band beam could be required. 

If the mass hierarchy is unknown the $\theta_{23}$ intrinsic degeneracy would be a part of a larger degeneracy similar to the conventional $\theta_{13} - \delta$ eightfold parameter degeneracy \cite{intrinsic,octant,MNjhep01}. Here, we focus on the ``intrinsic'' part by simply assuming that the mass hierarchy will already be determined at the time when the precision measurement we discuss will be realized. The generalization to the case including the unknown mass hierarchy can be performed in a straightforward manner.

This paper is organized as follows: in Sec. II we introduce the $\theta_{23}$ intrinsic degeneracy, the stimultaneous measurements of  $\delta$ and $\theta_{23}$ is discussed in detail in Sec. III \& IV. In Sec. V \& VI a simple toy analysis is given for a variety of possible future precision experiments. The conclusions are presented in Sec. VII.  The three Appendices contain some mathematical derivations too detailed for the main body of the paper.

\section{Introduction to the $\theta_{23}$ Intrinsic Degeneracy}
\label{intrinsic}

\begin{figure}[htbp]
\vspace{2mm}
\begin{center}
\includegraphics[width=0.46\textwidth]{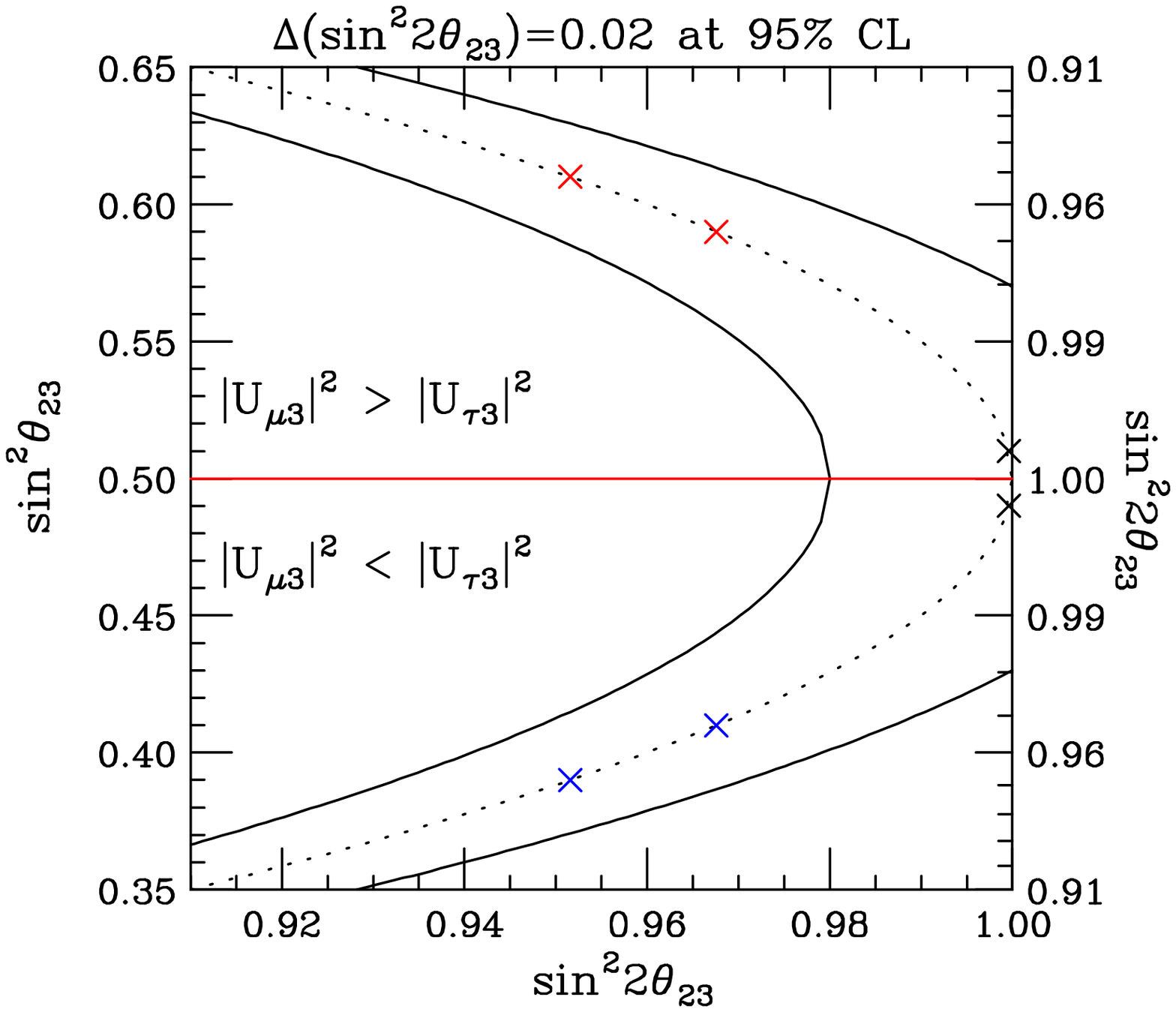}
\includegraphics[width=0.4\textwidth]{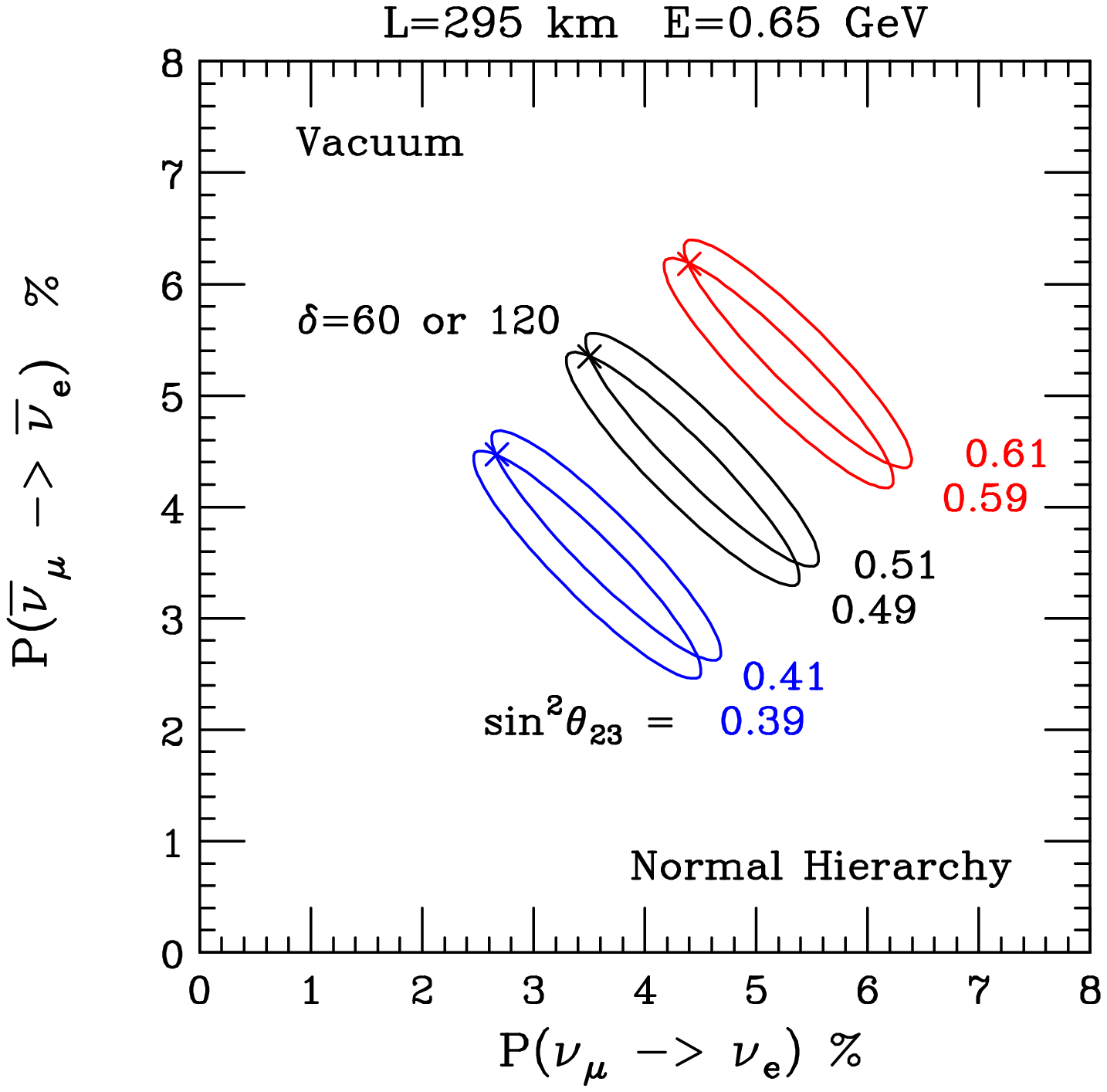}
\vspace{-3mm}
\end{center}
\caption{Left panel: Allowed region in $\sin^2 2\theta_{23}$ vs. $\sin^2 \theta_{23}$ space assuming 2\% uncertainty of $\sin^2 2\theta_{23}$ by a $\nu_\mu$ disappearance measurement. 
Right panel: The bi-probability plot in $P(\nu_{\mu} \to \nu_e)$ vs. $P(\bar{\nu}_{\mu} \to \bar{\nu}_e)$ space with three set of values of $P(\nu_{\mu} \to \nu_e)$ and $P(\bar{\nu}_{\mu} \to \bar{\nu}_e)$ which correspond to $\sin^2 2\theta_{23}$ approximately 0.96 (blue and red), 1.0 (black). The blue and red pair of ellipses are in different octants. The precise value of $\sin^2 \theta_{23}$ of each ellipse is as indicated in the figure, 
and at the crossing points indicated by $\times$ the value of $\delta$ is $60^\circ$ ($120^\circ$) for larger (smaller) $\sin^2 \theta_{23}$ ellipses,  for  the normal hierarchy.
The colored crosses on the left panel correspond to the ellipse on the right panel. }
\label{theta23-bi-P}
\end{figure}

We start by giving pedagogical reminder of the relevant issues, some of which have been discussed in the past, but not all of them. Suppose one could measure $\sin^2 2\theta_{23}$ to an uncertainty of 0.02 at, say, 95\% C.L. Then, to what accuracy can one determine $\sin^2 \theta_{23} \equiv s^2_{23}$? The answer to this question is given in the left panel in Fig.~\ref{theta23-bi-P}. In the region of $\theta_{23}$ far away from maximal, the loss of sensitivity to $s^2_{23}$ is modest, apart from the octant issue, whereas for $\sin^2 2\theta_{23} \gsim 0.98$ the accuracy of $s^2_{23}$ suddenly jumps to 10-20\% level. This was observed in  Fig.~3 of ref. \cite{MSS04} where they plotted the $s^2_{23}$-error as a function of $\sin^2 2\theta_{23}$. 

In the left panel in Fig.~\ref{theta23-bi-P}, two vertical scales are provided for convenience of the readers. Observe that the nonlinearity of scales in mapping from $s^2_{23}$ to $\sin^2 2\theta_{23}$ is large, and it is most significant in region where
\begin{eqnarray}
\sin^2 2\theta_{23} \gsim 0.97 \quad {\rm or}  \quad 0.41 \leq \sin^2 \theta_{23} \leq 0.59 \quad {\rm or} \quad 40^\circ \leq \theta_{23} \leq 50^\circ.
\label{theta23-range}
\end{eqnarray}
This is the region with significant overlap with the experimentally preferred one \cite{SK-atm,MINOS-disapp}, and it is the region 
we are primarily interested in for the purposes of this paper.

In the right panel of Fig.~\ref{theta23-bi-P}, the bi-probability plot  \cite{MNjhep01}, gives the allowed values of $(P(\nu_{\mu} \to \nu_e), ~P(\bar{\nu}_{\mu} \to \bar{\nu}_e))$ for three set of values of
 $\sin^2 2\theta_{23}=$0.9516, 0.9676 and 0.9996. Here, we have switched off the matter effect and restricted ourselves to the Normal Hierarchy. The figure shows that with an {\it exact} measurement of the $\nu_e$ and $\bar{\nu}_e$ appearance probabilities one can determine $s^2_{23}$ and $\delta$ at the same time apart from the twofold degeneracy, which we call the $\theta_{23}$ intrinsic degeneracy. Notice that this degeneracy is completely different in nature from the $\theta_{23}$ octant degeneracy and exists for all values of $\theta_{23}$, even at $\pi/4$. 

To illuminate the nature of the new degeneracy on $\theta_{23}$ and $\delta$, we place in the left panel of Fig.~\ref{theta23-bi-P} three pairs of cross marks, distinguished also by color, corresponding to three sets of ellipse shown in the bi-probability plot, the right panel.
In the two cases far off maximal mixing $\theta_{23}$ depicted by the blue and the red crosses, it is obvious that the degeneracy has nothing to do with the $\theta_{23}$ octant degeneracy. 
In the case near the maximal $\theta_{23}$ drawn by the black crosses in Fig.~\ref{theta23-bi-P}, the two solutions live in different octants, and we have a merging of the $\theta_{23}$ intrinsic degeneracy and the octant degeneracy for $\theta_{23}$ at very close to the maximal mixing.\footnote{
If the precise value of $\theta_{13}$ is known, adding only disappearance measurement produces the $\theta_{23}$ octant degeneracy. Adding only $\nu$ and $\bar{\nu}$ appearance measurement produces the $\theta_{23}$ intrinsic degeneracy, whose solutions may exist across the octant  boundary for $\theta_{23}$ very close to maximal mixing (black line and crosses in Fig.~\ref{theta23-bi-P}). With a finite error of $\theta_{13}$, the two degeneracies fuse in this region. 
}

We note that the size of the ellipses does not change much with different values of $\theta_{23}$ because they scale as $\propto \sin 2\theta_{23}$ which changes very little near $\theta_{23}=\pi/4$. If the value of $\theta_{13}$ had been changed instead of $\theta_{23}$, not only does the location of the bi-probability ellipse change but also its size changes with increasing or decreasing $\sin \theta_{13}$.
Note also that in vacuum with arbitrary values of $\Delta \equiv \frac{\Delta m^2_{32} L }{4E}$, and in matter at vacuum oscillation maxima (VOM, $\Delta = \left( \frac{\pi}{2} + n \right) \pi$), the $\theta_{23}$ intrinsic degeneracy is a $\delta$, $\pi-\delta$ degeneracy. It is because the difference between the degenerate solutions of $\sin \delta$ vanishes in vacuum (see (\ref{difference})), and only $\sin \delta$ is determined at VOM. 
This is reminiscent of the $\theta_{13}$ intrinsic degeneracy first discussed in \cite{intrinsic} but there are important differences.  Understanding the nature of this degeneracy will be elaborated in the following three sections, \ref{23-delta-vom}, \ref{23-delta-general}, and \ref{features}.

From the bi-probability plot given in Fig.~\ref{theta23-bi-P}, one can infer an interesting feature of the $\delta$ dependence of the error of $\delta$. The uncertainty of $\theta_{23}$, which allows the ellipses moving up and down, results in larger errors for $\delta$ determination at the both ends of the ellipses, i.e., around $\delta \sim \pm \frac{\pi}{2}$, than at $\delta \sim 0$, or $\pi$. Therefore, the large uncertainty of $s^2_{23}$ described in (\ref{theta23-range}) can produce broad peaks at $\delta \simeq \pm \frac{\pi}{2}$ in the $\delta$ dependence of $\delta$ error. In fact, this feature is seen in some of the figures (e.g., in Fig.~9) given in \cite{Pilar-etal}. 

\section{Correlated Measurements of $\theta_{23}$ and $\delta$ by $\nu_e$ and $\bar{\nu}_e$  Appearance Channels} 
\label{23-delta-vom}

We demonstrate our new strategy by showing that simultaneous determination of $\theta_{23}$ and $\delta$ is the way to proceed. To illuminate the point we consider a counting experiment to measure 
the $\nu_e$ and $\bar{\nu}_e$ appearance probabilities at a given energy in matter;
\begin{eqnarray}
P \equiv P(\nu_{\mu} \to \nu_e) &=& 2 s^2_{23} A_{\oplus}^2 + 
2 \epsilon \sin 2\theta_{23} A_{\oplus} A_{\odot} \cos \left(  \delta+ \Delta \right) + 2  c^2_{23} \epsilon^2 A_{\odot}^2 
\nonumber \\
\bar{P} \equiv P(\bar{\nu}_{\mu} \to \bar{\nu}_e) &=& 2 s^2_{23} \bar{A}_{\oplus}^2 + 
2 \epsilon \sin 2\theta_{23} \bar{A}_{\oplus} A_{\odot} \cos \left(  \delta- \Delta \right) + 2  c^2_{23} \epsilon^2 A_{\odot}^2 
\label{Pemu-matter}
\end{eqnarray}
where $\Delta \equiv \frac{ \Delta m^2_{31} L}{4E}$ and $\epsilon \equiv \frac{ \Delta m^2_{21} } { \Delta m^2_{31} } \simeq 0.03$. 
The $A$ functions in (\ref{Pemu-matter}) are defined as 
\begin{eqnarray}
A_{\oplus} &\equiv& 
\sqrt{2} s_{13} c_{13}  \left( \frac{ \Delta m^2_{31} }{ \Delta m^2_{31} - a } \right) 
\sin \left(\frac{ ( \Delta m^2_{31} - a) L}{4E}  \right) 
\nonumber \\
\bar{A}_{\oplus} &\equiv& 
\sqrt{2} s_{13} c_{13} \left( \frac{ \Delta m^2_{31} }{ \Delta m^2_{31} + a } \right) 
\sin \left(\frac{ ( \Delta m^2_{31} + a) L}{4E}  \right) 
\nonumber \\
A_{\odot} &\equiv& 
\sqrt{2} c_{12} s_{12} c_{13}
\left( \frac{ \Delta m^2_{31} } { a } \right) 
\sin \left(\frac{ a L}{4E}  \right) = \bar{A}_{\odot}
\label{A-def}
\end{eqnarray}
where
\begin{eqnarray}
\frac{ a }{ \Delta m^2_{31} } & = & { 2\sqrt{2} G_F N_e E_\nu \over \Delta m^2_{31}} \nonumber \\
&=&
8.5 \times 10^{-2} 
\left(\frac{\rho}{2.8 \ \mathrm{g/cm}^3}\right)
\left(\frac{Y_e}{0.5}\right)
\left(\frac{2.5 \times 10^{-3}\ \mathrm{eV}^2}{\Delta m^2_{31}}\right)
\left(\frac{E}{1 \ \mathrm{GeV}}\right).~
\label{vac-condition}
\end{eqnarray}
$A_{\odot}$ is of order unity in most regions of relevant experimental parameters, but $A_{\oplus}$ and $\bar{A}_{\oplus}$ are suppressed by smallness of $s_{13}$. 
The $A$'s are defined such that the atmospheric and the solar oscillation probabilities at the maximal mixing angle $\theta_{23} = \frac{\pi}{4}$,
$P_{\oplus}$ and $P_{\odot}$ respectively, are given by $P_{\oplus} = A_{\oplus}^2$ and $P_{\odot} = \epsilon^2 A_{\odot}^2$. 
For simplicity we will ignore the $2 c^2_{23}  \epsilon^2 A_{\odot}^2$ terms, in Eq.~(\ref{Pemu-matter}), in the rest of this paper as they are significantly smaller than the other terms.\footnote{
To include the effect of $2 \epsilon^2 c^2_{23} A_{\odot}^2$ terms, rewrite $c^2_{23}=1-s^2_{23}$ and then the $2 s^2_{23}  \epsilon^2 A_{\odot}^2$ can be included in the $A_\oplus^2$ terms and the $2 \epsilon^2 A_{\odot}^2$ can be moved to the left hand side.
}

It is instructive to quote an estimation of the ratio of $A_\oplus$ to $\epsilon A_\odot$ in near vacuum environment, $\frac{ a }{ \Delta m^2_{31} } \ll 1$, assuming the experiment is performed near the first oscillation maximum, $\Delta \sim 1$,
\begin{eqnarray}
\frac{A_{\oplus}}{ \epsilon A_{\odot}} 
&=& 
\left( \frac{ 2 s_{13} }{ \sin 2\theta_{12} } \right)
\frac{ \sin \Delta}{ \Delta } 
\left[ 1 + 
\frac{ a }{ \Delta m^2_{31} } \left( 1 - \Delta \cot \Delta \right)
\right] 
\left( \frac{\Delta m^2_{31}}{\Delta m^2_{21}} \right) \simeq 6.
\label{A-ratio2}
\end{eqnarray}
As we will show later, the size of this ratio is significant for the relative size of the uncertainties for the measurement of $\sin^2 \theta_{23}$ and $\delta$.

\subsection{Measuring $\theta_{23}$ and $\delta$ at the first vacuum oscillation maximum}
\label{1stVOM}

To show the point let us consider measurement at the first vacuum oscillation maximum (1st VOM), $\Delta = \frac{\pi}{2}$. Then, $\cos \left( \delta \pm \Delta \right) = \mp \sin \delta$. One can eliminate $\delta$ by combining the two equations in (\ref{Pemu-matter}), then
\begin{eqnarray} 
(s^2_{23})_0
&=&
\frac{ 1 }{ 2 \left( A_{\oplus} + \bar{A}_{\oplus} \right)} ~ \frac{1}{\sqrt{A_\oplus \bar{A}_\oplus}} 
\left[
\bar{P} \left( \frac{ A_{\oplus} }{ \bar{A}_{\oplus} } \right)^\frac{1}{2} 
+
P \left( \frac{ \bar{A}_{\oplus} }{ A_{\oplus} } \right)^\frac{1}{2}
\right].
\label{s23-0}
\end{eqnarray}
If instead we eliminate the first terms of (\ref{Pemu-matter}) in favor of $\sin \delta$ we obtain, 
\begin{eqnarray} 
\sin \delta_0
&=&
\frac{ 1 }{ 2  \sin 2\theta_{23}   \left( A_{\oplus} + \bar{A}_{\oplus} \right)}  \frac{1}{\epsilon A_{\odot}}
\left[
\bar{P} \left( \frac{ A_{\oplus} }{ \bar{A}_{\oplus} } \right) - 
P \left( \frac{ \bar{A}_{\oplus} }{ A_{\oplus} } \right)
\right].
\label{sin-delta-0}
\end{eqnarray}
$\cos \delta_0$ can be obtained as $\cos \delta_0 = \pm \sqrt{ 1 - \sin^2 \delta_0 }$. Notice that the sign ambiguity $\delta \leftrightarrow \pi - \delta$ cannot be resolved by measurement here.  Apart from the different linear combinations of $P$ and $\bar{P}$, these two equations differ by the overall factors of $1/\sqrt{A_\oplus \bar{A}_\oplus}$ and $1/\epsilon A_{\odot}$ in $s^2_{23}$ and $\sin \delta$, respectively.
%


We use the matter perturbation theory to obtain a simpler expression which allows intuitive understanding. We assume $\frac{a}{\Delta m^2_{31}} \ll 1$, which is an excellent approximation for T2K,
then we obtain, to first order in $\frac{a}{\Delta m^2_{31}}$, 
\begin{eqnarray} 
(s^2_{23})_0 &=&
\frac{ 1 }{ 8 s^2_{13} } 
\left[ \left( \bar{P} + P \right) + 
\frac{a}{\Delta m^2_{31}} 
\left( \bar{P} - P \right)
\right],
\nonumber \\
\sin \delta_0 &=& 
\frac{ 1 }{ 8 \epsilon J_r  \pi } 
\left[
(\bar{P} - P) + \frac{ 2 a }{ \Delta m^2_{31} } 
(\bar{P} + P) 
\right], 
\label{s23-sin-delta}
\end{eqnarray}
where $J_r \equiv c_{12} s_{12} c_{23} s_{23} s_{13} c^2_{13}$.  
Therefore, at VOM and in near vacuum environment $s^2_{23}$ is determined by sum of the probabilities $\bar{P} + P$, while $\sin \delta$ is governed by the difference $\bar{P} - P$. 
$\bar{P} + P$ and $\bar{P} - P$ switch their positions to induce sub-leading corrections due to matter effect into $s^2_{23}$ and $\sin \delta$, as shown in (\ref{s23-sin-delta}).

\subsection{Error of $s^2_{23}$ is smaller than that of $\sin \delta$}
\label{error-23-delta}

Let us estimate the expected errors in measurement of $s^2_{23}$ and $\sin \delta$, assuming that the errors for $P$ and $\bar{P}$ are independent with each other. We also assume settings in which the matter effect remains subdominant and use notations $A_{\oplus}^{vac}$ for $A_{\oplus}=\bar{A}_{\oplus}$ in vacuum. Using (\ref{s23-0}) and (\ref{sin-delta-0}), the errors, $\Delta (s^2_{23})$ and $\Delta (\sin \delta)$, are given by
\begin{eqnarray} 
\Delta (s^2_{23}) &\approx& 
\frac{ 1 }{ 4( A_{\oplus}^{vac})^2} 
\sqrt{ (\Delta P)^2 + (\Delta \bar{P})^2}
\nonumber \\
\Delta (\sin \delta) &\approx&
\frac{ 1 }{ 4  \sin 2\theta_{23}  \vert A_{\oplus}^{vac} \vert (\epsilon A_{\odot} ) } 
\sqrt{ (\Delta P)^2 + (\Delta \bar{P})^2}
\label{s23-sin-delta-error}
\end{eqnarray}
Notice that the two errors are related with each other at the 1st VOM by
\begin{eqnarray} 
\Delta (s^2_{23}) \approx \left( \frac{\epsilon A_{\odot} }{ \vert A_{\oplus}^{vac} \vert } \right) ~ \Delta (\sin \delta) \simeq \frac{1}{6} \Delta (\sin \delta)
\label{s23-sin-delta-error2}
\end{eqnarray}
Therefore, once the experiment starts to measure $\sin \delta$, then $s^2_{23}$ can be determined with an uncertainty which is 6 times smaller.  Thus, if a precision measurement of $\sin \delta$ is made, then the precision on  $s^2_{23}$  becomes competitive with the precision from the disappearance measurement, and in fact far exceeds it at around the maximal mixing.

Notice that by using the appearance channels we are free from the $\theta_{23}$ octant degeneracy inherent in the disappearance channel.
The expression of errors of $s^2_{23}$ and $\delta$ in more general setting, off VOM and in matter, is given in Appendix~\ref{error-general}.

If one estimates the $s^2_{23}$ error by using the simulation done by the Hyper-K collaboration \cite{HK-LOI}, assuming 1.5 years of neutrino and 3.5 years of antineutrino runs with beam power of 1.65 MW, one finds find that the precision on  $s^2_{23}$ exceeds the precision from the disappearance measurement around $\delta = \frac{\pi}{2}$. Of course, a full stimulation is needed to confirm this result, which is beyond the scope of this paper.

\subsection{Effect of uncertainties of $\theta_{23}$ and $\theta_{13}$ on $\delta$}

In this subsection, we discuss relative importance between uncertainties of $\theta_{23}$ and $\theta_{13}$ on $\delta$. Our discussion here will be valid both on and off VOM. As already mentioned, the $\theta_{23}$ and the $\theta_{13}$ intrinsic degeneracies have many common properties. For example, in near vacuum environment, both the $\theta_{13}$ and the $\theta_{23}$ intrinsic degeneracies produce $\delta \leftrightarrow \pi - \delta$ ambiguity, and the merging of the true and clone allowed regions around $|\delta| = \frac{\pi}{2}$ results in enhanced errors of $\delta$. Now, one can show, by the similar argument as we have given at the end of Sec.~\ref{intrinsic}, that the uncertainties of $\theta_{13}$ can affect the sensitivity to $\delta$. Then, the natural question is; Which uncertainties are more important, the one of $\theta_{13}$, or of $\theta_{23}$?

We argue that one can make a rough estimate of the relative importance of $\theta_{13}$ and $\theta_{23}$ uncertainties on $\delta$ determination in the following way. Notice that both of the uncertainties produce shift of the ellipses, up and down, along the diagonal line in the bi-probability plot (see Fig.~\ref{theta23-bi-P}), and this movement results in uncertainty in $\delta$. The uncertainties on $\theta_{13}$ and $\theta_{23}$, $\Delta \theta_{13}$ and $\Delta \theta_{23}$, can produce the same amount of shift in the center of ellipse (the first term in the probabilities in (\ref{Pemu-matter})) if they satisfy the relationship 
\begin{eqnarray} 
\Delta \theta_{23} \vert_{equal} \approx 
\frac{ \Delta \theta_{13} }{ \sin \theta_{13} }
\label{equal-shift}
\end{eqnarray}
where we have used the approximation $\sin 2\theta_{23}=1$.
Using the current data of $\theta_{13}$, $\sin^2 2\theta_{23}=0.089 \pm 0.011$ \cite{Daya-Bay}, the uncertainty on $\theta_{13}$ is 0.55$^\circ$, thus using Eq.~(\ref{equal-shift}), we can obtain the equivalent shift in $\theta_{23} $, $\Delta \theta_{23} \vert_{equal} \approx 3.7^\circ$. This is comparable, or smaller than the current uncertainties on $\theta_{23}$ given by $0.391 \leq \sin^2 \theta_{23} \leq 0.691$ at 90\% CL \cite{SK-nu2012}, which implies $\Delta \theta_{23} \vert_{current} \approx 6.7^\circ$ ($3.8^\circ$) for upward (downward) error at 1$\sigma$ CL. Therefore, even in the current situation the uncertainty of $\theta_{23}$ can be larger by a factor of $\simeq 2$ than that due to the $\theta_{13}$ error. Since the error of $\theta_{13}$ is expected to become much smaller by the continuing $\theta_{13}$ measurements at the time when measurement of $\delta$ is performed, reducing the $\theta_{23}$ error is of crucial importance for an accurate measurement of $\delta$. 

\section{Simultaneous Measurement of $\theta_{23}$ and $\delta$: General Case}
\label{23-delta-general}

In this section we present solutions of $s^2_{23}$ and $\delta$ for a given set of measurement of $P$ and $\bar{P}$ at generic value of $\Delta$. Instead of giving the explicit solutions we describe an iterative procedure to obtain the solutions of the degeneracy equations. The explicit solutions of the degeneracy equations are given in Appendix~\ref{D-solution}. 

\subsection{Properties of the intrinsic $\theta_{23} - \delta$ degeneracy}

To display characteristic properties of the intrinsic $\theta_{23} - \delta$ degeneracy we derive approximate relationship between the degenerate solutions. We define differences between two degenerate solutions as 
$\Delta s^2_{23} \equiv (s^2_{23})_1 - (s^2_{23})_2$,
$\Delta (\sin \delta) \equiv \sin \delta_1 - \sin \delta_2$, and 
$\Delta (\cos \delta) \equiv \cos \delta_1 - \cos \delta_2$.
Using (\ref{Pemu-matter}) we obtain 
\begin{eqnarray} 
\Delta (\cos \delta) \cos \Delta 
&=& 
- \frac{ (\bar{A}_{\oplus} + A_{\oplus}) }{ 2  ~(\epsilon A_{\odot} )} 
~\frac{ \Delta s^2_{23}}{\sin 2\theta_{23} }, 
\nonumber \\
\Delta (\sin \delta) \sin \Delta 
&=&
- \frac{ (\bar{A}_{\oplus} - A_{\oplus}) }{ 2   ~(\epsilon A_{\odot}) } 
~\frac{ \Delta s^2_{23}}{\sin 2\theta_{23} },
\label{difference}
\end{eqnarray}
under the approximation that $\sin 2\theta_{23}$ for the two solution is nearly equal, which is in fact an excellent approximation. This feature allows us to derive a simple relation for $\Delta (\sin \delta)$ as in (\ref{difference}), which is different from the $\theta_{13}$ intrinsic degeneracy.

From (\ref{difference}) one can infer two important characteristic properties of the $\theta_{23} - \delta$ degeneracy. 

\begin{itemize}
\item The difference in $\sin \delta$ comes from the matter effect, and hence $\Delta (\sin \delta)$ has to vanish in vacuum and the two solutions exhibit the $\delta \leftrightarrow \pi-\delta$ degeneracy.

\item At any VOM, ($\cos \Delta = 0 $), $\Delta s^2_{23}$ has to vanish. It is because the ellipses have zero width at VOM and again the two solutions exhibit the $\delta \leftrightarrow \pi-\delta$ degeneracy.
\end{itemize}

\noindent
Therefore, a measurement near VOM can primarily determine only $\sin \delta$, and in that case $\delta \leftrightarrow \pi - \delta$ degeneracy remains. Spectral information around VOM, if available, can be used to break this degeneracy.

\subsection{Solution to the $\theta_{23} - \delta$ degeneracy equation}

We start from (\ref{s23-0}) and (\ref{sin-delta-0}) which we call the zeroth-order solutions. Notice that we keep value of $\Delta$ in $A_{\oplus}$ and $\bar{A}_{\oplus}$ generic here, though it has been set to $\frac{\pi}{2}$ when we have discussed the solutions at VOM in the previous section. Under this understanding, we denote the zeroth-order solutions in (\ref{s23-0}) and (\ref{sin-delta-0}) as $(s^2_{23})_0$ and $\delta_0$, respectively, in this section. 
By going back to the original equations (\ref{Pemu-matter}) and 
using the zeroth-order solutions (\ref{s23-0}) and (\ref{sin-delta-0}) one can derive the equations
\begin{eqnarray} 
&& s^2_{23} 
= (s^2_{23})_0 -  \sin 2\theta_{23}  
\frac{  2 \epsilon A_{\odot} }{ \left( A_{\oplus} + \bar{A}_{\oplus} \right)} 
\cos \Delta \cos \delta 
\label{s23-sol} \\[0.3cm]
&& 
\left( \bar{A}_{\oplus} + A_{\oplus} \right) \sin \Delta \sin \delta - \left( \bar{A}_{\oplus} - A_{\oplus} \right) \cos \Delta \cos \delta = 
\left( \bar{A}_{\oplus} + A_{\oplus} \right) \sin \delta_0
\label{sin-delta-sol}
\end{eqnarray}
which can be solved iteratively\footnote{
Start with $s^2_{23}=1/2$ and calculate $\sin \delta_0$ using Eq.~(\ref{sin-delta-0}). Obtain $\delta$ from the solutions of Eq.~(\ref{sin-delta-sol}) 
and then use Eq.~(\ref{s23-sol}) to obtain $s^2_{23}$ and iterate. Since $\sin \delta_0$ depends on $\sin 2\theta_{23}$, the change in $\delta$ will be small and the iteration will converge rapidly near maximal mixing.
} 
to find the solution at generic values of $\Delta$. 

The second equation in (\ref{sin-delta-sol}) can be solved to produce the two solutions 
\begin{eqnarray} 
\delta &=& \phi + 
\arcsin \left( \frac{ \left( \bar{A}_{\oplus} + A_{\oplus} \right) \sin \delta_0 }
{ \sqrt{ \bar{A}_{\oplus}^2 + A_{\oplus}^2 - 2 \bar{A}_{\oplus} A_{\oplus}  \cos 2 \Delta } } \right),
\nonumber \\
\delta &=& \phi + \pi - 
\arcsin \left( \frac{ \left( \bar{A}_{\oplus} + A_{\oplus} \right) \sin \delta_0 }
{ \sqrt{ \bar{A}_{\oplus}^2 + A_{\oplus}^2 - 2 \bar{A}_{\oplus} A_{\oplus}  \cos 2 \Delta } } \right)
\label{delta-solution}
\end{eqnarray}
where 
\begin{eqnarray} 
\cos \phi &\equiv& 
\frac{ \left( \bar{A}_{\oplus} + A_{\oplus} \right) \sin \Delta }
{ \sqrt{ \bar{A}_{\oplus}^2 + A_{\oplus}^2 - 2 \bar{A}_{\oplus} A_{\oplus}  \cos 2 \Delta } } ,
\nonumber \\
\sin \phi &\equiv& 
\frac{ \left( \bar{A}_{\oplus} - A_{\oplus} \right) \cos \Delta }
{ \sqrt{ \bar{A}_{\oplus}^2 + A_{\oplus}^2 - 2 \bar{A}_{\oplus} A_{\oplus}  \cos 2 \Delta } } .
\label{phi-def}
\end{eqnarray}
Note, at VOM $\phi=0$ and thus the two solutions exhibit the $\delta \leftrightarrow \pi-\delta$ degeneracy and they have the same value of $s^2_{23}$ since $\cos \Delta=0$.   In vacuum, Eq.~(\ref{sin-delta-sol}) reduces to  $\sin \delta= \sin \delta_0/\sin \Delta$, so that $\sin \delta$ is determined but $\cos \delta$ has two signs so that different values of $s^2_{23}$ are possible, provided that we are not at VOM.

\begin{figure}[htbp]
\vspace{2mm}
\begin{center}
\includegraphics[width=0.46\textwidth]{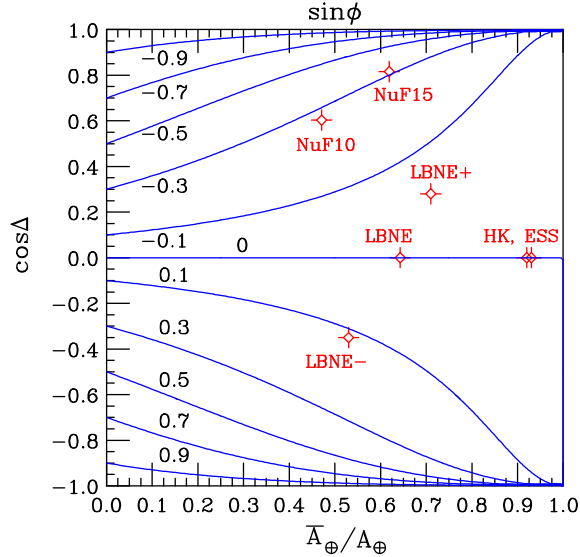}
\end{center}
\caption{Equi-$\sin \phi$ contours are plotted on $\frac{ \bar{A}_{\oplus} }{ A_{\oplus} }$ vs. $ \cos \Delta$ plane assuming the normal mass hierarchy. The typical values of $\frac{ \bar{A}_{\oplus} }{ A_{\oplus} }$ and $ \cos \Delta$ to be examined in Sec.~\ref{features} are also shown by asterisks; 
LBNE$\pm$ denote the cases with $20\%$ higher ($+$) and lower ($-$) energies than the one of the LBNE setting at VOM, and NuF10 and NuF15 stand for neutrino factory setting with energy 10 GeV and 15 GeV, respectively.}
\label{sinphi}
\end{figure}

We plot in Fig.~\ref{sinphi} the contours of equal $\sin \phi$ line on $\frac{ \bar{A}_{\oplus} }{ A_{\oplus} }$ vs. $\frac{\Delta}{\pi}$ plane, assuming the normal mass hierarchy for which $0 \leq \frac{ \bar{A}_{\oplus} }{ A_{\oplus} } \leq 1$. For the inverted mass hierarchy the abscissa must be interpreted as $\frac{ A_{\oplus} }{ \bar{A}_{\oplus} }$.\footnote{
In a very limited phase space $\frac{ \bar{A}_{\oplus} }{ A_{\oplus} }$ can be negative. But since it does not occur in most of the proposed settings we do not enter into this possibility here. 
} 
%
In Fig.~\ref{sinphi} we also show by asterisks the points to be examined in Sec.~\ref{features} with typical values of the various settings of experiments proposed. It is remarkable to see that in all cases $\sin \phi$ is small, despite one's naive expectation that might expect become large when matter effects are large e.g., in the LBNE and neutrino factory settings. 

\section{Features of simultaneous measurement of $\theta_{23}$ and $\delta$ in Some Experimental Settings}
\label{features}

In this section, we examine features of simultaneous measurement of $\theta_{23}$ and $\delta$ by considering several experimental settings with their typical parameters (baseline and energy). In the course of analysis, we illuminate some characteristic features of the $\theta_{23}$ intrinsic degeneracy. Though our analysis is based on assumed measurement of oscillation probabilities $P$ and $\bar{P}$ with errors, it may capture main features of the more extensive analysis with cross sections, efficiencies, energy spectra etc. with systematic errors. We hope to come back to such an analysis in the future. We start by describing our analysis method. 

\subsection{Analysis method}

We define the model $\chi^2$ which is used in our analysis as follows:
\begin{eqnarray}
\chi^2(\theta_{23},\delta) = \frac{(P(\theta_{23},\delta)-P(\theta^x_{23},\delta^x))^2}{(\Delta P)^2} + \frac{(\bar{P}(\theta_{23},\delta)-\bar{P}(\theta^x_{23},\delta^x))^2}{(\Delta \bar{P})^2}
\end{eqnarray}
where $\theta^x_{23}$ and $\delta^x$ are the test values for $\theta_{23}$ and $\delta$. In our analysis $\chi^2$ is computed only at some fixed values of neutrino energy and spectrum information is not taken into account.

We determine the errors to be placed on measurement of $P$ and $\bar{P}$ in the following way. The number of event $N$ in a single ``energy bin'' may be calculated as $N=n_{T} \int dE f (E) \frac{d \sigma }{d E} P \approx n_{T} f_0 \frac{d \sigma }{d E} P$, 
where $n_{T}$ denotes target number density and $f_0 = f (E) \Delta E$ is the neutrino flux in a particular energy bin of width $\Delta E$. Ignoring the systematic uncertainties on flux, cross sections, and target number density, and assuming the remaining uncertainties are of statistical nature only, then $\frac{\Delta N}{N} = \frac{\Delta P}{P} \propto 1/\sqrt{P}$ since $\Delta N$ varies as $\sqrt{N}$. 
Thus, we will assume the fractional uncertainty on the oscillation probabilities are given by
\begin{eqnarray} 
\frac{ \Delta P }{P} = 0.03 \sqrt{ \frac{P_0}{P} },
\hspace{8mm}
\frac{ \Delta \bar{P} }{ \bar{P} } = 0.03 \sqrt{ \frac{\bar{P}_0}{ \bar{P} } }.
\label{error-def}
\end{eqnarray}
The reference probabilities are taken as $P_0=\bar{P}_0=\sin^2 \theta_{23} \sin^2 2 \theta_{13}= 0.5 \times 0.09 = 0.045$, so that about 1100 events are assumed at the reference point, Equation~(\ref{error-def}) will be referred to as  ``3\% error'' throughout this section. 
(In Sec.~\ref{nufact}, we will use 1\% error just by replacing 0.03 by 0.01 in (\ref{error-def}) in a neutrino factory analysis.)

\subsection{Hyper-K vs. LBNE}
\label{Hyper-K}

\begin{figure}[htbp]
\vspace{4mm}
\begin{center}
\includegraphics[width=0.46\textwidth]{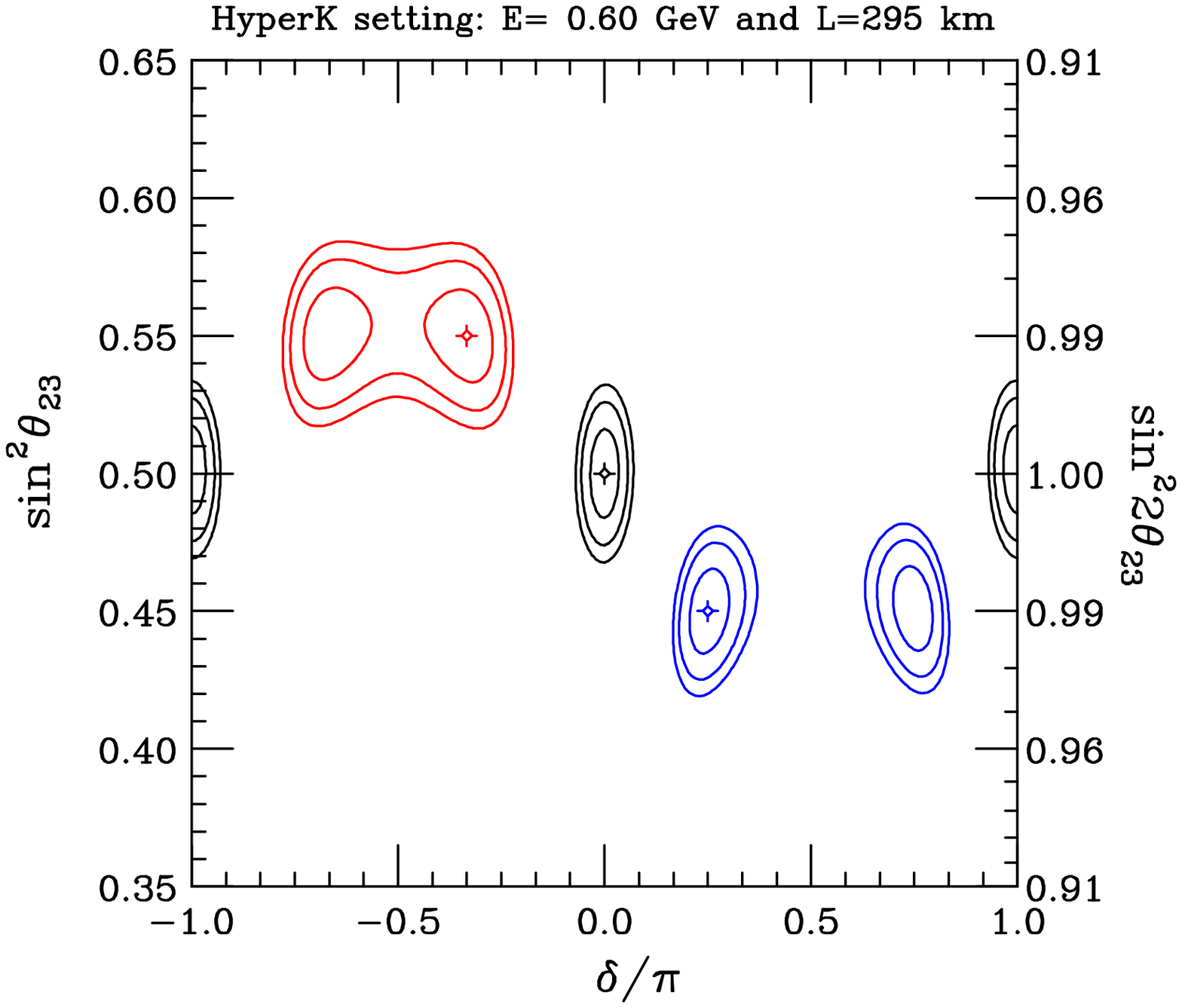}
\includegraphics[width=0.46\textwidth]{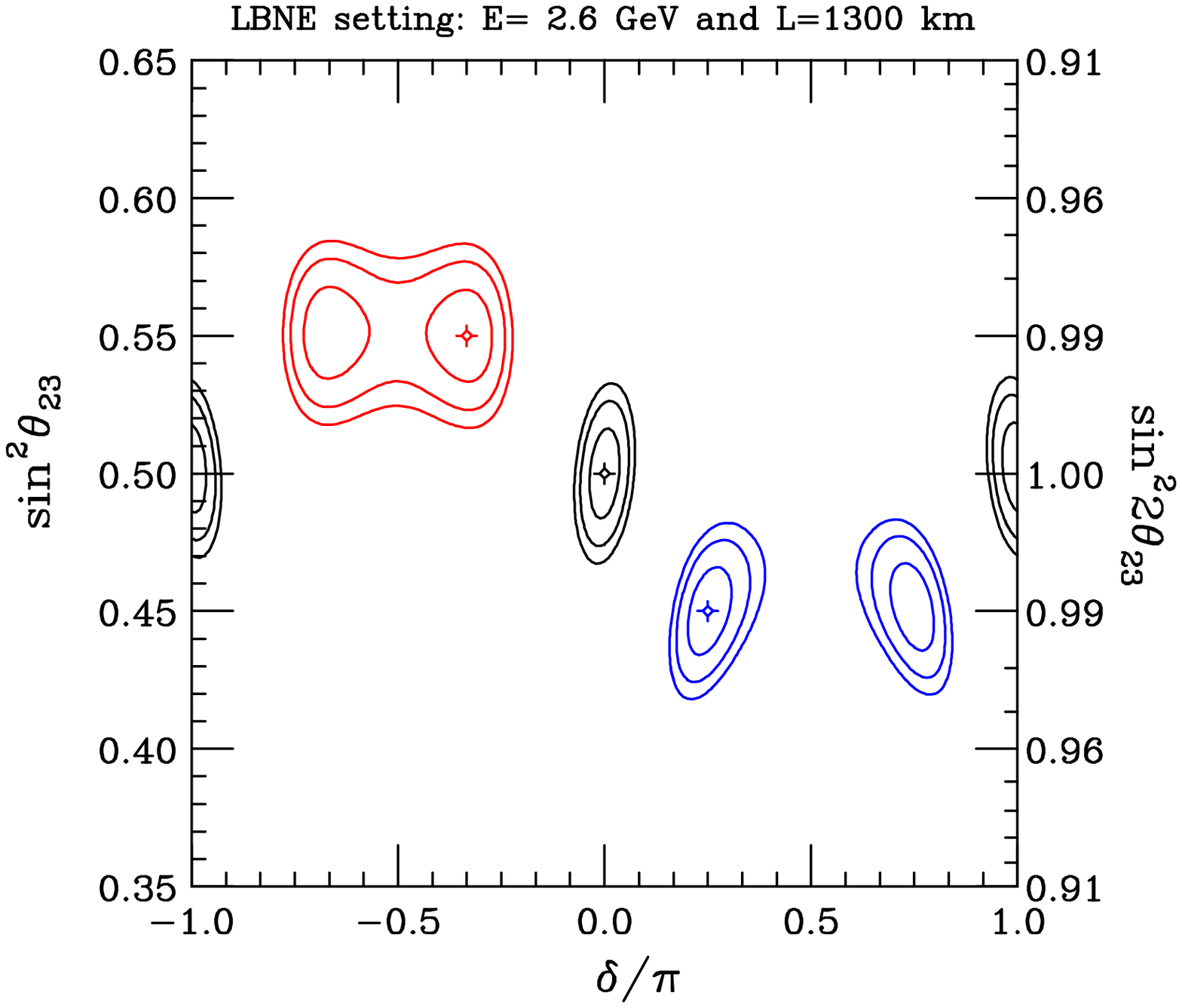}
\end{center}
\caption{Contours of allowed regions at $1\sigma$, $2\sigma$, and $3\sigma$ CL  (2 d.o.f.) in $\delta / \pi$ vs. $\sin^2 \theta_{23}$ space assuming 3\% error for both $P$ and $\bar{P}$ measurement. The left and right panels are for the Hyper-K and the LBNE settings, respectively. 
The true values of ($\delta$, $\sin^2 \theta_{23}$) are taken as ($0^\circ$, 0.5) the case shown in black, ($45^\circ$, 0.45) in blue, and ($- 60^\circ$ 0.55) in red.
}
\label{HK-LBNE-VOM}
\end{figure}

Let us start by examining the Hyper-K and the LBNE settings at the VOM, by which we mean $E=0.6 $ GeV and $L=295$ km for Hyper-K \cite{HK-LOI}, and $E=2.6$ GeV and $L=1300$ km for the LBNE settings \cite{LBNE}, respectively. In Fig.~\ref{HK-LBNE-VOM}, the allowed contours at $1\sigma$, $2\sigma$, and $3\sigma$ CL  (2 d.o.f.) in $\delta / \pi$ vs. $\sin^2 \theta_{23}$ space are presented in which a 3\% error is assumed for both $P$ and $\bar{P}$ measurement. Three sets of values of $\delta$ and $\sin^2 \theta_{23}$ are used as inputs: ($0^\circ$, 0.5) shown by black contour, ($45^\circ$, 0.45) by blue contours, and ($- 60^\circ$ 0.55) by red contours. They are shown by $+$ symbol in the figure, the style we keep throughout this section. 

%

The features of the allowed regions shown in the left and right panels of Fig.~\ref{HK-LBNE-VOM} are remarkably similar with each other despite the larger matter effect in the LBNE setting. One may tempted to interpret this feature as simply due to that they are both at the VOM. But, actually it is not the whole story as explained shortly below.

We observe in Fig.~\ref{HK-LBNE-VOM} for any one of the inputs two allowed regions corresponding to the true and the clone solutions, a manifestation of the $\theta_{23}$ intrinsic degeneracy. They are related by symmetry under $\delta \leftrightarrow \pi - \delta$ as it is clearly visible in the figure. The symmetry reflects the fact that the setting is almost at VOM at which only $\sin \delta$ survives in the oscillation probabilities. It is also notable that true and clone contours are horizontally spaced, $\Delta s^2_{23} \simeq 0$, the feature which also arises by sitting at VOM, as seen in (\ref{difference}). The allowed regions become larger as $\sin^2 \theta_{23}$ get larger because the errors of $P$ and $\bar{P}$ increases by 
$\approx 10 \%$ from $\sin^2 \theta_{23}=0.45$ (blue) to 0.55 (red). 

\begin{figure}[htbp]
\vspace{2mm}
\begin{center}
\includegraphics[width=0.46\textwidth]{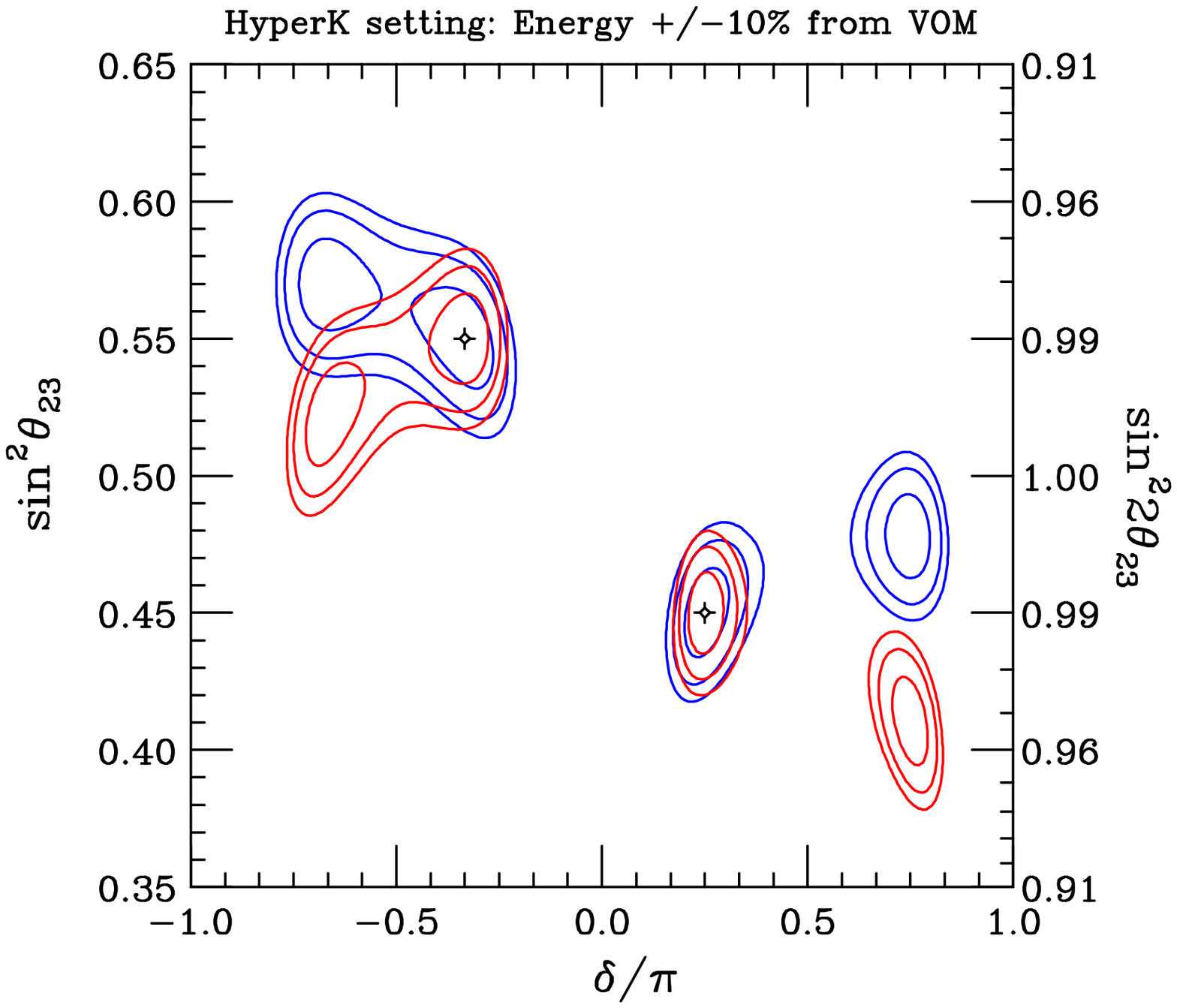}
\includegraphics[width=0.46\textwidth]{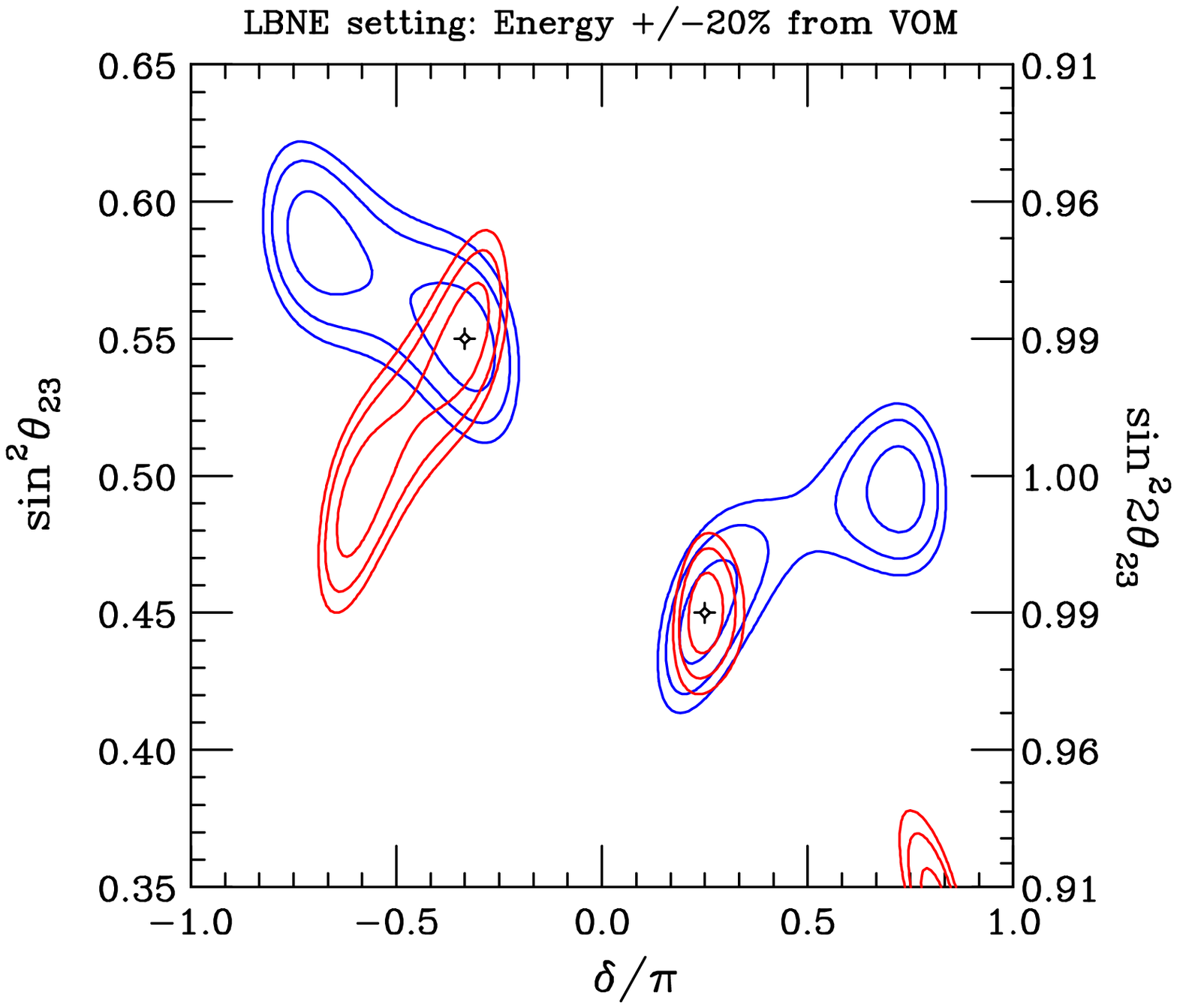}
\end{center}
\caption{The same as in Fig.~\ref{HK-LBNE-VOM} but with higher ($\Delta < \frac{\pi}{2}$, shown in blue) or lower ($\Delta > \frac{\pi}{2}$, shown in red) neutrino energies by $\pm 10\%$ (Hyper-K) and $\pm 20\%$ (LBNE) than the ones of VOM. The true values of ($\delta$, $\sin^2 \theta_{23}$) are taken as ($45^\circ$, 0.45) shown in blue, and ($- 60^\circ$ 0.55) shown in red.}
\label{HK-LBNE-Evary}
\end{figure}

In Fig.~\ref{HK-LBNE-Evary}, the similar contours of allowed regions are plotted with the same Hyper-K and LBNE settings, but with higher (shown in blue) and lower (shown in red) energies by $10 \%$ ($20 \%$) for Hyper-K (LBNE) than their respective energies at VOM. Our choice of larger variations of energy off VOM reflects the fact that LBNE is equipped with a wide-band neutrino beam.  
In fact, if we choose the same rate of variation of energy for Hyper-K and LBNE the features of the contours are almost the same. The matter effect plays little role for this plot. It is because the slope $\frac{ \Delta \cos \delta }{ \Delta s^2_{23}}$ in (\ref{difference}) is insensitive to the matter effect; it cancels out to leading order in $\frac{a}{\Delta m^2_{32}}$. We observe that the direction of shift of $\cos \delta$ is opposite as that of $s^2_{23}$ for $\Delta < \frac{\pi}{2}$ (shown in blue), while it goes along the same direction for $\Delta > \frac{\pi}{2}$ (shown in red), as it should be according to (\ref{difference}). 

In Fig.~\ref{HK-LBNE-Evary}, we observe that while the allowed regions for the true and the clone solutions remain separated at $1 \sigma$ CL (apart from the red one with input $s^2_{23} = 0.55$ for LBNE) they start to merge together at higher CL. Once the merging takes place the errors are governed primarily by distance between the true and the clone solutions, in addition to the uncertainties inherent to the measurement. If this phenomenon occurs in sizable fraction of the parameter space of the analysis the accuracies of measurement of both $s^2_{23}$ and $\sin \delta$ are significantly lost. To our knowledge, this feature does not appear to be addressed in previously performed sensitivity studies for $\delta$, while it might have been buried into the necessarily intricate simulation procedure.

We notice that the movement of the allowed regions are rather dynamic despite the small shift of the energies in both cases. It suggests that the spectrum informations above and below VOM in the off-axis (narrow band) beam would be important to resolve the $\theta_{23}$ intrinsic degeneracy. Naturally, the distances between the degenerate solutions become farther apart for larger energy variation as shown in Fig.~\ref{HK-LBNE-Evary}, suggesting that LBNE with the wide-band beam would be more powerful in resolving the $\theta_{23}$ intrinsic degeneracy.

\subsection{A neutrino factory setting}
\label{nufact}

\begin{figure}[htbp]
\vspace{2mm}
\begin{center}
\includegraphics[width=0.46\textwidth]{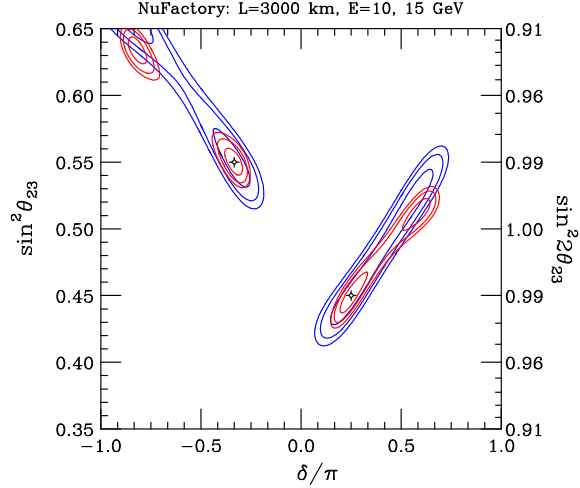}
\vspace{-2mm}
\end{center}
\caption{Contours of allowed regions at $1\sigma$, $2\sigma$, and $3\sigma$ CL  (2 d.o.f.) in $\delta / \pi$ vs. $\sin^2 \theta_{23}$ space for neutrino factory setting with $L=3000$ km. Two particular values of neutrino energies are taken for comparison, one at $E=10$ GeV (shown in red) and $E=15$ GeV (shown in blue). Unlike other cases 1\% error is assumed for both $P$ and $\bar{P}$ measurement. 
The true values of ($\delta$, $\sin^2 \theta_{23}$) are taken as ($45^\circ$, 0.45) and ($- 60^\circ$ 0.55). 
}
\label{NF_evary}
\end{figure}

In Fig.~\ref{NF_evary}, the similar allowed contours are presented for a particular setting for neutrino factory \cite{nufact} with assumed baseline of $L=3000$ km. Two values of neutrino energy, $E=10$ GeV and $E=15$ GeV, are chosen as typical ones which are shown in red and in blue, respectively. Unlike other cases we take 1\% errors for $P$ and $\bar{P}$, which implies $\simeq 10^4$ events in each channel at the reference probabilities $P_0=\bar{P}_0=0.045$. 
Despite the smaller errors taken discriminating power of the clone solution from the true one is quite limited. Two regions of true and clone solutions merge apart from those in region of $\sin \delta < 0$ at $E=10$ GeV. Since they are below VOM ($\Delta < \frac{\pi}{2}$) the relationship between true and clone solutions are similar at both energies.

\section{A setting at the second oscillation maximum}
\label{2ndVOM} 

In this section we discuss some new aspects that show up when the beam energy is tuned to, or it covers over, the second VOM, $\Delta = \frac{3 \pi}{2}$. The most important change in this region 
is that $A_{\odot} \propto \Delta$ becomes larger by a factor of three. Then, $\delta$ dependent term which comes from interference between $A_{\oplus}$ and $A_{\odot}$ gets larger by the same amount. It influences to the errors of $s^2_{23}$ and $\delta$. $\Delta s^2_{23}$ in (\ref{s23-sin-delta-error}) remains unchanged if the uncertainties $\Delta P$ and $\Delta \bar{P}$ are the same at the 2nd VOM. On the other hand, $\Delta (\sin \delta)$ becomes smaller by a factor of three, implying a better sensitivity to CP at the 2nd VOM. Therefore, Eq.~(\ref{s23-sin-delta-error2}) is modified to 
$\Delta (s^2_{23}) \simeq \frac{1}{2} \Delta (\sin \delta)$.

\begin{figure}[htbp]
\vspace{2mm}
\begin{center}
\includegraphics[width=0.46\textwidth]{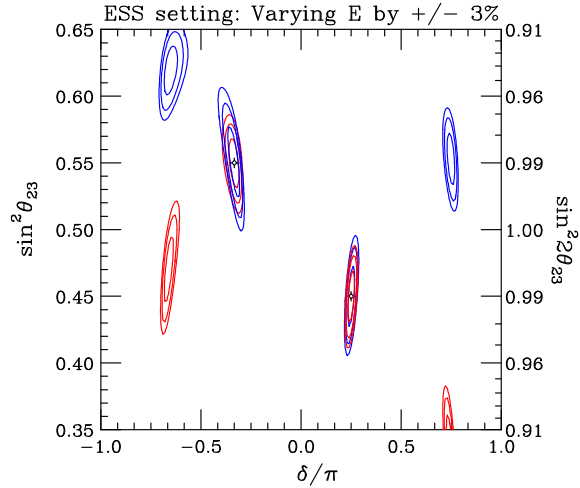}
\vspace{-2mm}
\end{center}
\caption{Contours of allowed regions at $1\sigma$, $2\sigma$, and $3\sigma$ CL  (2 d.o.f.) in $\delta / \pi$ vs. $\sin^2 \theta_{23}$ space for ESS setting with $L=540$ km. Two particular values of neutrino energies are taken for comparison, one at $E=0.342$ GeV (shown in red) and $E=0.364$ GeV (shown in blue), which amount to $\pm 3\%$ of the one at 2nd VOM. The true values of ($\delta$, $\sin^2 \theta_{23}$) are taken as ($45^\circ$, 0.45) and ($- 60^\circ$ 0.55). 
}
\label{ESS_evary}
\end{figure}

To elucidate the new features of the setting at 2nd VOM we examine the case with ESS (European Spallation Source) \cite{ESS}, the facility to be build in Sweden which plans to produce 5 MW of beam power which can be used to construct intense neutrino beam. The baseline to a detector is assumed to be $L=540$ km and the energy at around the 2nd VOM is given approximately as $E=0.353$ GeV. In Fig.~\ref{ESS_evary} presented are the contours of allowed regions by assuming 3\% measurement of $P$ and $\bar{P}$. We take neutrino energies 3\% higher (0.364 GeV, $\Delta < \frac{\pi}{2}$, shown in blue), or 3\% lower (0.342 GeV, $\Delta > \frac{\pi}{2}$, shown in red) than that of 2nd VOM. 

One notices that the allowed regions significantly shrink along $\delta$ direction, the effect we mentioned above. It is also remarkable that clone solutions moves significantly upward or downward despite very small variation of neutrino energy of $\pm$ 3\%. It implies that resolving power of the $\theta_{23}$ intrinsic degeneracy, in principle, can be higher at the 2nd VOM.

\section{Conclusion}

In this paper, we have shown that $s^2_{23}$ can be determined automatically at higher accuracies than $\sin \delta$ in experiments which can measure $\delta$ using $\nu_e$ and $\bar{\nu}_e$ appearance channels. The accuracy for $s^2_{23}$ determination supersedes that of the conventional method which relies on $\nu_\mu$ disappearance channels in region $40^\circ \lsim  \theta_{23} \lsim 50^\circ $, assuming high enough statistics that allows reasonable precision on the measurement of CP phase, $\delta$. The reasons for such high accuracy is that the appearance method is free from the problems of Jacobian broadening in the translation $\sin^2 2\theta_{23} \rightarrow s^2_{23}$ for values of  $\theta_{23}$ near $45^\circ$, and of the determination of the octant  of $\theta_{23}$.  

With use of $\nu_e$ and $\bar{\nu}_e$ appearance channels, determination of $\theta_{23}$ is inherently coupled with measurement of CP phase $\delta$. 
Thus, the uncertainties on the measurement of $\theta_{23}$ and $\delta$ are correlated. Furthermore, our treatment reveals  that measurement of the two independent parameters metamorphoses into a new strategy which may be called as ``simultaneous measurement of $\theta_{23}$ and $\delta$''. Reflecting on the nature of these measurements, a new parameter degeneracy called the ``$\theta_{23}$ intrinsic degeneracy'' is identified. In general, two degenerate solutions of $\theta_{23}$ and $\delta$ are given for each set of measurements of oscillation probabilities, $P$ and $\bar{P}$, producing a new ambiguity and a new source of uncertainties, hitherto unrecognized. This degeneracy must be resolved to make an accurate determination of both $\theta_{23}$ and $\delta$. 

To understand characteristic features of correlated measurement of $\theta_{23}$ and $\delta$, we have undertaken the following two approaches: 
(1) Toward resolution of the degeneracy, we have analyzed the natures of the $\theta_{23}$ intrinsic degeneracy, and obtained the analytic formulas for the degeneracy solutions which help us to understand characteristic features of the degeneracy. 
(2) We have simulated simple data set of $P$ and $\bar{P}$ with suitably assumed errors to obtain the contours of allowed regions in $s^2_{23}$ vs. $\delta$ space. We have examined various settings, Hyper-K, LBNE, and neutrino factory all around (or below) 1st VOM, and ESS (European Spallation Source) near 2nd VOM, using their typical values of baseline and neutrino energies. 

From these studies we have observed the following features of our new method, simultaneous measurement of $\theta_{23}$ and $\delta$. The obtained contours of allowed regions confirm our expectation that the error of $s^2_{23}$ determination is approximately six times smaller than that of $\sin \delta$ apart from the problem of merging the degeneracy solutions. Importantly, the movement of clone solutions against varying energy is generally quite large, and the effect is more significant around 2nd VOM than the 1st VOM. Thus, we have uncovered the crucial point that spectrum information is of key importance to resolve the $\theta_{23}$ intrinsic degeneracy. This feature is reminiscent of the one encountered in the $\theta_{13}$ intrinsic degeneracy. 

In addition, the contours of allowed regions revealed some characteristic features of the $\theta_{23}$ intrinsic degeneracy, such as $\delta \rightarrow \pi - \delta$ symmetry, in most of the settings we discussed. 
While the feature must hold at VOM or in vacuum, its robustness in fact requires explanation. We have shown that it can be understood by our analytical expressions of the degeneracy solutions, which implies that the matter effect plays only a minor role in the relationship between the two solutions of $\delta$.

We have concluded that our new method, using $\nu_e$ appearance channels, offers a better way of measuring $\theta_{23}$ with high accuracy, in the region $40^\circ \lsim \theta_{23} \lsim 50^\circ$ in particular, and the simultaneous measurement of $\theta_{23}$ and $\delta$ is the right strategy to carry out precision measurement of the both parameters. Incorporation of these ideas into a full stimulation will be undertaken in a follow up paper \cite{CMP}.

\newpage
\appendix

\section{Uncertainty of $\delta$ from Probabilities and $s^2_{23}$} 
\label{delta-error}

In this Appendix, we calculate the uncertainties in $\cos \delta$ and $\sin \delta$ which would be produced by small measurement uncertainties in $P$, $\bar{P}$, and $s^2_{23}$. Here, we do {\em not} consider $s^2_{23}$ as a parameter determined by our appearance method with  $(P ,\bar{P})$ measurement, but regard it as an external parameter determined e.g., by using the disappearance channel.

If we start with the appearance probability expressions, Eq.~(\ref{Pemu-matter}), and expand the $\cos(\Delta \pm \delta)$ terms: there are two terms with $\delta$ dependence, one proportional to $\sin \delta$ and the other proportional to $\cos \delta$. 
We then use the chain rule of differentiation, to leading order in $\epsilon \equiv \frac{ \Delta m^2_{21} }{ \Delta m^2_{31} }$ and for $\cos 2 \theta_{23} \ll 1$, to obtain\footnote{
Eqs.~(\ref{cos-sin-delta-uncertainty}) and (\ref{cos-sin-delta-uncertainty2}) is more accurate if you interpret $P$ and $\bar{P}$ in the denominator in the first two terms of these equations as $\langle P \rangle$ and $\langle \bar{P} \rangle$, the ones averaged over $\delta$. In other words $\langle P \rangle$ and $\langle \bar{P} \rangle$ imply the coordinate of the center of the bi-P ellipses.
}
%
\begin{eqnarray}
\hspace*{-0.6cm}
\left. \begin{array}{l}
\Delta (\cos \delta \cos \Delta) \\[0.2cm]
\Delta (\sin \delta \sin \Delta)
\end{array} \right\}
&=& 
\left( \frac{ \tan \theta_{23} }{4 } \right) 
\left[ 
\left( \frac{ \bar{A}_{\oplus} }{ \epsilon  A_{\odot} } \right)
 \frac{\Delta \bar{P}}{ \bar{P} } 
\pm
\left( \frac{ A_{\oplus} }{ \epsilon  A_{\odot} } \right)
\frac{\Delta P}{ P } 
+ 
\left( \frac{ \bar{A}_{\oplus} \pm A_{\oplus} }{ \epsilon  A_{\odot} } \right) \frac{ \Delta (s^2_{23}) }{ s^2_{23}}
\right] ~~~~~
\label{cos-sin-delta-uncertainty}
\end{eqnarray}
We interpret this result as follows:
the uncertainties on the measurement of $(\sin \delta \sin \Delta)$ or of $(\cos \delta \cos \Delta)$, assuming that the uncertainties on $P$, $\bar{P}$, and $s^2_{23}$ are uncorrelated,  are given by 
\begin{eqnarray} 
\hspace*{-0.3cm}
\left. \begin{array}{l}
\Delta (\cos \delta \cos \Delta) \\[0.2cm]
\Delta (\sin \delta \sin \Delta)
\end{array} \right\} & =&
\left( \frac{ \tan \theta_{23} }{4  } \right) 
\sqrt{ 
\left( \frac{ \bar{A}_{\oplus} }{ \epsilon A_{\odot} } \right)^2 (\mathcal{D} \bar{P})^2 +  \left( \frac{ A_{\oplus} }{ \epsilon A_{\odot} } \right)^2 (\mathcal{D} P)^2  + 
 \left( \frac{ \bar{A}_{\oplus}\pm A_{\oplus} }{ \epsilon A_{\odot} }   \right)^2 (\mathcal{D} s^2_{23})^2
}
\nonumber \\
& & \label{cos-sin-delta-uncertainty2}
\end{eqnarray}
where $\mathcal{D} O \equiv \frac{\Delta O}{O}$ denotes fractional error of the quantity $O$.

In vacuum, when $A_{\oplus} = \bar{A}_{\oplus}$, it's clear that uncertainty on $\sin \delta$ does not depend on the uncertainty of $s^2_{23}$ whereas the uncertainty on $\cos \delta$ depends strongly on this uncertainty.  In matter, however, the uncertainty on both  $\sin \delta$ and $\cos \delta$  will depend on the uncertainty on $s^2_{23}$.  One needs both measurements to make a precise measurement of $\delta$
otherwise the measurement will suffer from a degeneracy similar to the $\delta \leftrightarrow \pi-\delta$ degeneracy seen at VOM. In the Introduction we simplified this rather complicate situation to make a point regarding the importance of an accurate determination of $s^2_{23}$.

In near vacuum environment, $\frac{ a }{ \Delta m^2_{31} } \ll 1$, $\bar{A}_{\oplus} \approx A_{\oplus}$ and 
we have 
$\frac{ \tan \theta_{23} }{4} \frac{ A_{\oplus} }{  \epsilon  A_{\odot} } \simeq \frac{6}{4 } \simeq 1.5$, which entails the coefficients in $\sqrt{ \sum_{X} (\delta X)^2}$ in (\ref{cos-sin-delta-uncertainty2}) of order unity. Therefore, in near vacuum, an  estimation of the uncertainties on $(\sin \delta \sin \Delta)$ (or $(\cos \delta \cos \Delta)$) are given by a very simple expressions 
\begin{eqnarray} 
\text{Uncertainty of~} (\sin \delta \sin \Delta)
& \sim & f_{\text{error}} \sqrt{ (\mathcal{D} P)^2 + (\mathcal{D} \bar{P})^2 }, \label{sin-delta-error-final} \\
\text{Uncertainty of~} (\cos \delta \cos \Delta)
& \sim  &f_{\text{error}} \sqrt{ (\mathcal{D} P)^2 + (\mathcal{D} \bar{P})^2 + 4 (\mathcal{D} s^2_{23})^2 }
\label{cos-delta-error-final}
\end{eqnarray}
where $f_{\text{error}}$ is a coefficient of order unity.\footnote{
In matter there exists $\eta (\mathcal{D} s^2_{23})^2$ term in the square root in (\ref{sin-delta-error-final}) but with small coefficient. For example, $\eta \simeq 0.24$ for LBNE.
}
%
Thus, for $|\sin \delta | $ near $1$, where a precise measurement of $\cos \delta$ is important for an accurate determination of the angle $\delta$, the uncertainty in $s^2_{23}$ plays a significant role. 
The feature that $\sin \delta$ does not depend on $s^2_{23}$ while $\cos \delta$ does in the near vacuum environment can be easily understood by the bi-probability plot given in the right panel of Fig.~\ref{theta23-bi-P}. When $s^2_{23}$ is varied within the uncertainty ``center of gravity'' of the ellipse moves along the diagonal line in $P - \bar{P}$ plane, which affects predominantly $\cos \delta$ but not so much to $\sin \delta$, as this is a measure of the distance away from the diagonal. 

The results in (\ref{sin-delta-error-final}) and (\ref{cos-delta-error-final}) are consistent with each other because here $\sin \delta$ and $\cos \delta$ are treated as independent.  Otherwise, the relationship $\Delta (\cos \delta) = - \tan \delta \Delta (\sin \delta)$ must hold. 

\section{Errors of $s^2_{23}$ and $\delta$ off the VOM}
\label{error-general}

One may ask the question; Can one derive the formula of errors of $s^2_{23}$ and $\delta$ analogous to (\ref{s23-sin-delta-error}) in general setting off VOM and without vacuum dominated approximation? In this Appendix we give an answer to this question.

We start from the expressions of $P$ and $\bar{P}$ in (\ref{Pemu-matter}). We denote the changes of $P$ and $\bar{P}$ induced by a small changes of $s^2_{23}$ and $\delta$ as $\Delta P$ and $\Delta \bar{P}$. Conversely, it may interpreted as the relations between small shifts of $s^2_{23}$ and $\delta$ and their cause, i.e., errors of $P$ and $\bar{P}$. To leading order in the small variations we obtain the relationship between the shifts $\Delta s^2_{23}$, $\Delta \delta$, $\Delta P$ and $\Delta \bar{P}$ as
\begin{eqnarray} 
\Delta s^2_{23} 
& \approx & 
\frac{ 
\left[  \Delta \bar{P} ~\left( \frac{A_{\oplus}} {\bar{A}_{\oplus}} \right)^\frac{1}{2} \sin(\Delta+\delta) + \Delta P ~\left( \frac{\bar{A}_{\oplus}}{A_{\oplus} } \right)^\frac{1}{2}  \sin(\Delta-\delta)
\right] }
{ 2 \sqrt{\bar{A}_{\oplus} A_{\oplus} }
\left[ \left( \bar{A}_{\oplus} + A_{\oplus} \right) \sin \Delta \cos \delta + 
 \left(  \bar{A}_{\oplus} - A_{\oplus}  \right) \cos \Delta \sin \delta 
 \right] },
 \label{s23-error-general}
\end{eqnarray}
\begin{eqnarray} 
\Delta \delta 
&\approx &
\frac{ 
\left[  \Delta \bar{P} ~\left(\frac{ A_{\oplus} }{ \bar{A}_{\oplus} }\right) -  \Delta P ~\left(\frac{ \bar{A}_{\oplus} }{ A_{\oplus} } \right) \right] }
{ 2 \sin 2\theta_{23} (\epsilon A_{\odot})
\left[ \left( \bar{A}_{\oplus} + A_{\oplus}  \right) \sin \Delta \cos \delta + 
 \left(  \bar{A}_{\oplus} - A_{\oplus}  \right) \cos \Delta \sin \delta 
 \right] },
\label{delta-error-general}
\end{eqnarray}
where we have ignored the variation of $\sin 2 \theta_{23}$ to derive these expressions, since we are only interested in the region near maximal mixing.

From these expressions it is straightforward to derive the formulas of errors of $s^2_{23}$ and $\delta$ by assuming that the errors of $P$ and $\bar{P}$ are independent with each other, and by ignoring errors in neutrino energies, $\Delta m^2_{32}$, and etc. They generalize (\ref{s23-sin-delta-error}) to the case in matter and off VOM. Notice that these expressions of errors are obtained in our appearance method for simultaneous measurement of $\theta_{23}$ and $\delta$, and they should not be confused with the errors discussed in Appendix~\ref{delta-error}.

\section{Solution of the Degeneracy Equation}
\label{D-solution}

Here, we obtain explicit solutions to the degeneracy problem involving $s_{23}$ and $\delta$.

\subsection{General solutions within the normal hierarchy}

By using (\ref{Pemu-matter}) we obtain 
\begin{eqnarray}
\cos \Delta \cos \delta 
&=& \frac{1}{4 \epsilon \sin 2\theta_{23} A_{\odot} } 
\left[ \left( \frac{ \bar{P} }{ \bar{A}_{\oplus} } + \frac{ P }{ A_{\oplus} } \right) - 
2 s^2_{23} \left( \bar{A}_{\oplus} + A_{\oplus} \right) 
\right],
\nonumber \\ 
\sin \Delta \sin \delta &=& \frac{1}{4 \epsilon \sin 2\theta_{23} A_{\odot} } 
\left[ \left( \frac{ \bar{P} }{ \bar{A}_{\oplus} } - \frac{ P }{ A_{\oplus} } \right) - 
2 s^2_{23} \left( \bar{A}_{\oplus} - A_{\oplus} \right) 
\right].
\label{cos-sin-delta}
\end{eqnarray}
Then, the equality $\cos^2 \delta + \sin^2 \delta =1$ can be written in a form of quadratic equation for $x \equiv s^2_{23}$:
\begin{eqnarray} 
a x^2 - b x + c = 0
\label{x-equation}
\end{eqnarray}
where 
\begin{eqnarray} 
a &=& \left( \bar{A}_{\oplus}^2 + A_{\oplus}^2 \right) - 2 \bar{A}_{\oplus} A_{\oplus} \cos 2 \Delta + 4 \epsilon^2 A_{\odot}^2 \sin^2 2 \Delta
\nonumber \\
b &=& \left(  \bar{P} + P \right) - \left( \frac{ \bar{P} }{ \bar{A}_{\oplus} } A_{\oplus} + \frac{ P }{ A_{\oplus} } \bar{A}_{\oplus} \right) \cos 2 \Delta 
+ 4 \epsilon^2 A_{\odot}^2 \sin^2 2 \Delta
\nonumber \\
c &=& \frac{1}{4}
\left[ 
\left( \frac{ \bar{P} }{ \bar{A}_{\oplus} } \right)^2 + \left( \frac{ P }{ A_{\oplus} }  \right)^2 - 
2 \left( \frac{ \bar{P} }{ \bar{A}_{\oplus} } \right) \left( \frac{ P }{ A_{\oplus} }  \right) \cos 2 \Delta 
\right]
\label{abc}
\end{eqnarray}
The solution to (\ref{x-equation}) is of course given by  
\begin{eqnarray} 
x = \frac{ b \pm \sqrt{b^2 - 4 ac} }{ 2a }
\label{x-solution}
\end{eqnarray}
The solutions for $\cos \delta$ and $\sin \delta$ can be obtained by inserting (\ref{x-solution}) into (\ref{cos-sin-delta}).
The discriminant $D \equiv b^2 - 4 ac$ is given by 
\begin{eqnarray} 
D &=& \sin^2 2 \Delta 
\biggl[
- \left( \frac{ \bar{P} }{ \bar{A}_{\oplus} } A_{\oplus} - \frac{ P }{ A_{\oplus} } \bar{A}_{\oplus} \right)^2 
\nonumber \\
&+&
4 \epsilon^2 A_{\odot}^2 \biggl\{
2 (\bar{P} + P) - \left( \frac{ \bar{P} }{ \bar{A}_{\oplus} }^2 + \frac{ P }{ A_{\oplus} }^2 \right)  - 2 \cos 2 \Delta \left( \frac{ \bar{P} }{ \bar{A}_{\oplus} } A_{\oplus} + \frac{ P }{ A_{\oplus} } \bar{A}_{\oplus} -  \frac{ \bar{P} }{ \bar{A}_{\oplus} } \frac{ P }{ A_{\oplus} } \right) 
\biggr\} 
\nonumber \\
&+& 16 \epsilon^4 A_{\odot}^4 \sin^2 2 \Delta
\biggr]
\label{x-discriminant}
\end{eqnarray}
When $D > 0$ (\ref{x-solution}) describes two solutions of $s_{23}$ and $\delta$ for a given set of observable $P$ and $\bar{P}$. By inserting them into (\ref{cos-sin-delta}) one obtains the corresponding solutions of $\cos \delta$ and $\sin \delta$.

The degeneracy solutions within the alternative (flipped $\Delta$-sign) mass hierarchy can be obtained by making replacements $\Delta \rightarrow - \Delta$, $A_{\oplus} \rightarrow - \bar{A}_{\oplus} $, and $\bar{A}_{\oplus} \rightarrow - A_{\oplus}$ in (\ref{x-solution}) and (\ref{abc}).

\subsection{Perturbation around the zeroth-order solution}

We obtain perturbative solution of $s^2_{23} \equiv x$ and $\delta$ in a form $x = x_0 + x_1$, and $\delta=\delta_0 + \delta_1$ to leading order of small deviation of $\Delta = \frac{\pi}{2} - y$. Here, $x_0$ and $\delta_0$ imply the zeroth-order solution given in (\ref{s23-0}) and (\ref{sin-delta-0}).\footnote{
Strictly speaking, since we do not take $\Delta = \frac{\pi}{2}$ in $A_{\odot}$ and $\bar{A}_{\odot}$ in zeroth-order solutions it is not quite the perturbative expansion. Yet, it is a useful way to obtain approximate solutions near the VOM. It may be justified when $\bar{A}_{\oplus}$ and $A_{\oplus}$ are slowly varying around VOM. When we really talk about the solution at VOM, $A_{\odot}$ and $\bar{A}_{\odot}$ must be interpreted as taking $\Delta = \frac{\pi}{2}$ in them. 
}
%
Notice that $\cos \Delta \approx y$, $\cos 2 \Delta \approx -1 + \frac{y^2}{2}$, and $\sin 2 \Delta \approx 2 y$.
We obtain the equation for $x_1$:
\begin{eqnarray}
&&\left(\bar{A}_{\oplus} + A_{\oplus} \right)^2 x_1^2 
\nonumber \\
&=& 
4 \epsilon^2 A_{\odot}^2 \sin 2\theta_{23} y^2 + 
4 \bar{A}_{\oplus} A_{\oplus} x_0^2 y^2 - 
2 \left( \frac{ \bar{P} }{ \bar{A}_{\oplus} } A_{\oplus} + \frac{ P }{ A_{\oplus} } \bar{A}_{\oplus} \right) x_0 y^2 + 
\frac{ \bar{P} }{ \bar{A}_{\oplus} } \frac{ P }{ A_{\oplus} } y^2
\label{x-perturb}
\end{eqnarray}
where a small changes to $\sin 2\theta_{23}$ due to correction to $x_0$ is ignored. 
The solution to this equation is given by 
\begin{eqnarray}
x_1 = \pm \frac{y}{\left(\bar{A}_{\oplus} + A_{\oplus} \right)} 
2 \epsilon A_{\odot} \sin 2\theta_{23} \cos \delta_0
\label{x-perturb-solution}
\end{eqnarray}
where we have used (\ref{sin-delta-0}). Using $\sin (\delta_0 + \delta_1) \approx \sin \delta_0 +  \delta_1 \cos \delta_0 $ in (\ref{cos-sin-delta}) one obtains, ignoring order $y^2$ term, 
\begin{eqnarray}
\delta_1 &=& - \frac{\bar{A}_{\oplus} - A_{\oplus}}{ 2 \epsilon A_{\odot} \sin 2\theta_{23} } x_1 =
\mp y \frac{\bar{A}_{\oplus} - A_{\oplus}}{ \bar{A}_{\oplus} + A_{\oplus} } \cos \delta_0 
\label{delta-perturb-solution}
\end{eqnarray}
where we have used (\ref{x-perturb-solution}). The correction to $\delta$ comes from the matter effect, whereas the one to $x=s^2_{23}$ is dominantly vacuum effect.

\begin{acknowledgments}

We thank Pilar Coloma for useful discussions. 
H.M. thanks Theory Group of Fermilab for warm hospitality during visit in winter in 2012-2013 where this work was started and essentially completed. He is grateful to CNPq for support which enables him to visit to Dept.~F\'{\i}sica, PUC in Rio de Janeiro. 
He is also supported in part by KAKENHI received through TMU, Grant-in-Aid for Scientific Research No. 23540315, Japan Society for the Promotion of Science.

S.P.  acknowledges partial support from the  European Union FP7  ITN INVISIBLES (Marie Curie Actions, PITN- GA-2011- 289442).
Fermilab is operated by the Fermi Research Alliance under contract no. DE-AC02-07CH11359 with the U.S. Department of Energy.

\end{acknowledgments}


\begin{thebibliography}{99}

\bibitem{Daya-Bay} 
F.~P.~An {\it et al.}  [DAYA-BAY Collaboration],
  Phys.\ Rev.\ Lett.\  {\bf 108}, 171803 (2012)
  [arXiv:1203.1669 [hep-ex]]; 
  arXiv:1210.6327 [hep-ex].

\bibitem{RENO} 
J.~K.~Ahn {\it et al.}  [RENO Collaboration],
  Phys.\ Rev.\ Lett.\  {\bf 108}, 191802 (2012)
  [arXiv:1204.0626 [hep-ex]].

\bibitem{DChooz} 
Y.~Abe {\it et al.}  [DOUBLE-CHOOZ Collaboration],
  Phys.\ Rev.\ Lett.\  {\bf 108}, 131801 (2012)
  [arXiv:1112.6353 [hep-ex]]; 
  Phys.\ Rev.\ D {\bf 86}, 052008 (2012)
  [arXiv:1207.6632 [hep-ex]].

\bibitem{T2K} 
  K.~Abe {\it et al.}  [T2K Collaboration],
  Phys.\ Rev.\ Lett.\  {\bf 107}, 041801 (2011)
  [arXiv:1106.2822 [hep-ex]].

\bibitem{MINOS} 
P.~Adamson {\it et al.}  [MINOS Collaboration],
  Phys.\ Rev.\ Lett.\  {\bf 107}, 181802 (2011)
  [arXiv:1108.0015 [hep-ex]]; 
  [arXiv:1301.4581 [hep-ex]].
  
  \bibitem{SK-atm} 
  R.~Wendell {\it et al.}  [Super-Kamiokande Collaboration],
  Phys.\ Rev.\ D {\bf 81}, 092004 (2010)
  [arXiv:1002.3471 [hep-ex]].

\bibitem{MINOS-disapp} 
P.~Adamson {\it et al.}  [MINOS Collaboration],
  Phys.\ Rev.\ Lett.\  {\bf 106}, 181801 (2011)
  [arXiv:1103.0340 [hep-ex]].

\bibitem{MSS04}
  H.~Minakata, M.~Sonoyama and H.~Sugiyama,
  Phys.\ Rev.\ D {\bf 70}, 113012 (2004) 
  [hep-ph/0406073].

\bibitem{intrinsic}
  J.~Burguet-Castell, M.~B.~Gavela, J.~J.~Gomez-Cadenas, P.~Hernandez and O.~Mena,
  Nucl.\ Phys.\ B {\bf 608}, 301 (2001) 
  [hep-ph/0103258].

\bibitem{octant}
G.~L.~Fogli and E.~Lisi,
Phys.\ Rev.\ D {\bf 54}, 3667 (1996)
[arXiv:hep-ph/9604415].

\bibitem{MNjhep01}
H.~Minakata and H.~Nunokawa,
  JHEP {\bf 0110}, 001 (2001)
  [arXiv:hep-ph/0108085].

\bibitem{Pilar-etal}
P.~Coloma, A.~Donini, E.~Fernandez-Martinez and P.~Hernandez,
  JHEP {\bf 1206}, 073 (2012)
  [arXiv:1203.5651 [hep-ph]].

\bibitem{HK-LOI}
  K.~Abe, 
  {\it et al.}, ``Letter of Intent: The Hyper-Kamiokande Experiment --- Detector Design and Physics Potential ---,''
  arXiv:1109.3262 [hep-ex].

\bibitem{SK-nu2012}
Y.~Itow, Talk at XXIV International Conference on Neutrino Physics and Astrophysics (Neutrino 2012), Kyoto, Japan, June 3-9, 2012. 

\bibitem{LBNE}
  T.~Akiri {\it et al.}  [LBNE Collaboration],
  arXiv:1110.6249 [hep-ex].

\bibitem{nufact}
A.~Bandyopadhyay {\it et al.}  [ISS Physics Working Group Collaboration],
  Rept.\ Prog.\ Phys.\  {\bf 72}, 106201 (2009)
  [arXiv:0710.4947 [hep-ph]].

\bibitem{ESS}
E.~Baussan, M.~Dracos, T.~Ekelof, E.~F.~Martinez, H.~Ohman and N.~Vassilopoulos,
  arXiv:1212.5048 [hep-ex].

\bibitem{CMP}
P.~Coloma, H.~Minakata, S.~Parke, in preparation.

\end{thebibliography}
\end{document}